\documentclass[aps,pra,noshowpacs,twocolumn,letterpaper,superscriptaddress,longbibliography,floatfix]{revtex4-1}

\usepackage{graphicx}
\usepackage{amsmath,amssymb,amsfonts}
\usepackage{braket}
\usepackage{mathtools}
\usepackage{color}

\graphicspath{{./figs/},{./figs_suppl/}}

\DeclareMathOperator{\Tr}{Tr}

\begin{document}

\title{All-Optical Storage of Phase-Sensitive Quantum States of Light}
\author{Yosuke~Hashimoto}
\affiliation{Department of Applied Physics, School of Engineering, The University of Tokyo, 7-3-1 Hongo, Bunkyo-ku, Tokyo 113-8656, Japan}
\author{Takeshi~Toyama}
\affiliation{Department of Applied Physics, School of Engineering, The University of Tokyo, 7-3-1 Hongo, Bunkyo-ku, Tokyo 113-8656, Japan}
\author{Jun-ichi~Yoshikawa}
\email{yoshikawa@ap.t.u-tokyo.ac.jp}
\affiliation{Department of Applied Physics, School of Engineering, The University of Tokyo, 7-3-1 Hongo, Bunkyo-ku, Tokyo 113-8656, Japan}
\author{Kenzo~Makino}
\affiliation{Department of Applied Physics, School of Engineering, The University of Tokyo, 7-3-1 Hongo, Bunkyo-ku, Tokyo 113-8656, Japan}
\author{Fumiya~Okamoto}
\affiliation{Department of Applied Physics, School of Engineering, The University of Tokyo, 7-3-1 Hongo, Bunkyo-ku, Tokyo 113-8656, Japan}
\author{Rei~Sakakibara}
\affiliation{Department of Applied Physics, School of Engineering, The University of Tokyo, 7-3-1 Hongo, Bunkyo-ku, Tokyo 113-8656, Japan}
\author{Shuntaro~Takeda}
\affiliation{Department of Applied Physics, School of Engineering, The University of Tokyo, 7-3-1 Hongo, Bunkyo-ku, Tokyo 113-8656, Japan}
\author{Peter~van~Loock}
\affiliation{Institute of Physics, Johannes Gutenberg-Universit\"{a}t Mainz, Staudingerweg 7,
55099 Mainz, Germany}
\author{Akira~Furusawa}
\email{akiraf@ap.t.u-tokyo.ac.jp}
\affiliation{Department of Applied Physics, School of Engineering, The University of Tokyo, 7-3-1 Hongo, Bunkyo-ku, Tokyo 113-8656, Japan}

\date{\today}

\begin{abstract}
We  experimentally demonstrate storage and on-demand release of phase-sensitive, photon-number superposition states of the form $\alpha \ket{0} + \beta e^{i\theta} \ket{1}$ for an optical quantized oscillator mode.
For this purpose, we introduce a phase-probing mechanism to a storage system composed of two concatenated optical cavities, which was previously employed for storage of phase-insensitive single-photon states [Phys.\ Rev.\ X \textbf{3}, 041028 (2013)].
This is the first demonstration of all-optically storing highly nonclassical and phase-sensitive quantum states of light.
The strong nonclassicality of the states after storage becomes manifest as a negative region in the corresponding Wigner function shifted away from the origin in phase space. 
This negativity is otherwise, without the phase information of the memory system, unobtainable. While our scheme includes the possibility of optical storage, on-demand release and synchronization of arbitrary single-rail qubit states, it is not limited to such states. 
In fact, our technique is extendible to more general phase-sensitive states such as multiphoton superposition or entangled states, and thus it represents a significant step toward advanced optical quantum information processing, where highly non-classical states are utilized as resources.
\end{abstract}


\maketitle

Optical quantum states are indispensable as flying quantum information carriers, and furthermore, they can serve as sufficient resources for universal and fault-tolerant quantum computing.
Prominent examples of photonic quantum information processing are the Knill-Laflamme-Milburn (KLM) scheme \cite{Knill.Nature2001}, based on discrete variables and photon detections, and the Gottesman-Kitaev-Preskill (GKP) scheme \cite{Gottesman.PRA2001,Fukui.PRL2017}, based on continuous variables and homodyne detections. 
In these applications, quantum memories of optical states are necessary.
More specifically, a frequently executed process like gate teleportation \cite{Gottesman.Nature1999,Bartlett.PRL2003}, by which a probabilistically induced nonlinearity can be efficiently incorporated into a quantum protocol,
relies, to a certain extent, on quantum memories in order to synchronize the probabilistically generated auxiliary resource states.

Even though many experiments with quantum memories have been performed for a couple of decades \cite{Lvovsky.NPhoton2009}, only recently optical single-photon states with high purity, having a negative region in the Wigner function, were stored and retrieved on demand \cite{Yoshikawa.PRX2013,Bimbard.PRL2014}.
Storage of highly pure single-photon states was demonstrated there with two different memory systems, one based on concatenated cavities \cite{Yoshikawa.PRX2013} and the other based on atomic ensembles \cite{Bimbard.PRL2014}.
The cavity systems were then used for synchronization of two photons showing Hong-Ou-Mandel interference \cite{Makino.SciAdv2016}.

The above high-purity memories are, in principle, capable of storing arbitrary quantum states of a harmonic oscillator $\sum_{n=0}^\infty c_n\ket{n}$, where photon-number eigenstates $\ket{n}$ form an orthonormal basis.
However, for the successful demonstration of such a universal quantum memory, there has been a technical challenge, that is, phase stabilization of the optical quantum states to be stored.
The Fock states $\ket{n}$, including the single-photon state $\ket{1}$, and their incoherent mixtures, are phase-insensitive states and thus highly pure single-photon states have been stored in previous demonstrations without tracking the optical phase inside the memory.
On the other hand, coherent superpositions of Fock states are phase-sensitive.
If one loses the phase information, off-diagonal elements of the density matrix $\ket{n}\bra{m}$ with $n\neq m$ are effectively diminished, which finally results in mixed phase-insensitive states $\sum_{n=0}^\infty \rho_{n,n}\ket{n}\bra{n}$.
With such phase-unpreserving memories, many important resource states lose their quantum features, such as cubic phase states (i.e., the ancillae for the GKP scheme), 
superpositions of coherent states \cite{Ralph.PRA2003}, and entangled states in the case of two or more modes.

In this Letter, we demonstrate storage and controlled release of general phase-sensitive superpositions of vacuum and one-photon states $c_0\ket{0}+c_1\ket{1}$ ($c_0, c_1\in\mathbb{C}$) for a single optical mode, i.e., an, in principle, arbitrary single-rail qubit.
We introduce a phase reference to the concatenated cavity system \cite{Yoshikawa.PRX2013,Makino.SciAdv2016}, and also a mechanism for creating phase-sensitive superposition states.
As a result, we demonstrate for the first time all-optical storage of phase-sensitive quantum states with a negative region of the Wigner function shifted from the origin in phase space.
This negative region originates from the off-diagonal elements $\ket{0}\bra{1}$, $\ket{1}\bra{0}$ of the density matrix, and it is not preserved without the phase information.
The negative region is preserved for a storage time of up to about 200 ns, but the genuine non-Gaussianity (any non-Gaussian features indescribable by incoherently mixing Gaussian states) is kept for even longer, namely for about 400 ns \cite{supplemental}.
Beyond storage of arbitrary single-rail (single-mode) qubits, even more importantly, our phase-preserving memory scheme is further applicable to arbitrary optical quantum states $\ket{\psi} = \sum_{n} c_n \ket{n}$, and paves the way for synchronization of phase-sensitive states, which is fundamentally important for both optical quantum computing \cite{Takeda.Nature2013,Takeda.PRL2017} and quantum communication \cite{Duan.Nature2001}.

Creation of highly pure non-Gaussian states typically require strong nonlinearities, which is still difficult to obtain deterministically at present.
Therefore, we resort to probabilistic schemes, where photon detections herald successful events.
We start from two-mode squeezed vacuum states $\ket{\Psi} \propto \sum_{n \geq 0} (\tanh \gamma)^n \ket{n}_\text{s} \ket{n}_\text{i}$ obtained by parametric down conversion, where signal (s) and idler (i) fields are entangled.
Then we perform measurements on the idler field.
If the idler field is subject to a photon counter and the measurement outcome is one photon, the signal field is projected onto a single-photon state.
A more general measurement on the idler field, introducing auxiliary coherent beams, in principle enables one to project the signal field onto arbitrary states of a single mode \cite{Bimbard.NPhoton2010,Yukawa.OptExpress2013}, as well as certain classes of multi-mode states \cite{Yoshikawa.PRA2018}.
However, the times of successful events in the heralding scheme are random.
Memory systems are powerful, because they give timing controllability to probabilistically prepared states.

\begin{figure}[tbp]
\centering
\includegraphics[width = 0.45\textwidth]{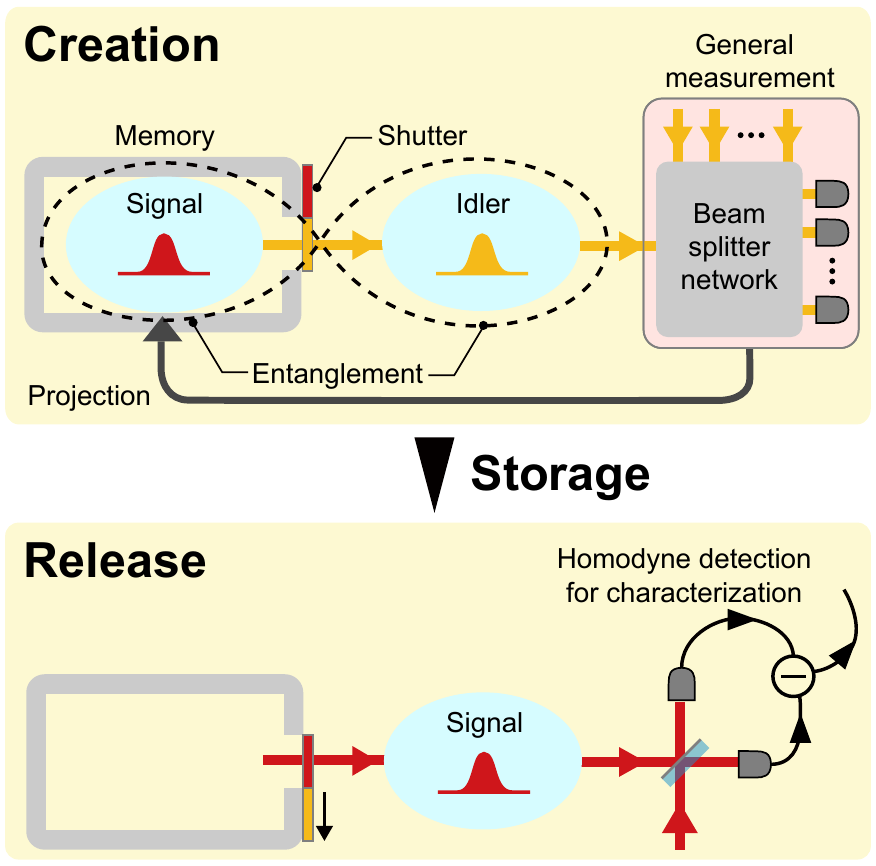}
\caption{Conceptual diagram of creation, storage, and on-demand release of phase-sensitive quantum states.
The memory and shutter are implemented by concatenated cavities.
}
\label{fig:sequence}
\end{figure}

The concept of our memory system to store and release general superposition states is shown in Fig.~\ref{fig:sequence}.
Parametric down conversion occurs inside a memory cavity, where the signal and idler fields are separated in frequency by the free spectral range of the memory cavity \cite{Yoshikawa.PRX2013}.
The shutter at the output of the memory is actually another cavity, which can transmit either the signal or the idler via the resonance condition.
Initially, the shutter cavity transmits the idler field, whereas the signal field is not transmitted.
By performing general measurements on the emerging idler field, the signal field inside the memory cavity is projected onto various states.
Whenever the created signal state should be available for further processing and exploitation, it can be released at a controlled timing by shifting the resonance of the shutter cavity.
The resonance shift is done by utilizing an electro-optic modulator (EOM) contained inside the shutter cavity.

\begin{figure*}
\centering
\includegraphics[width = 0.9\textwidth]{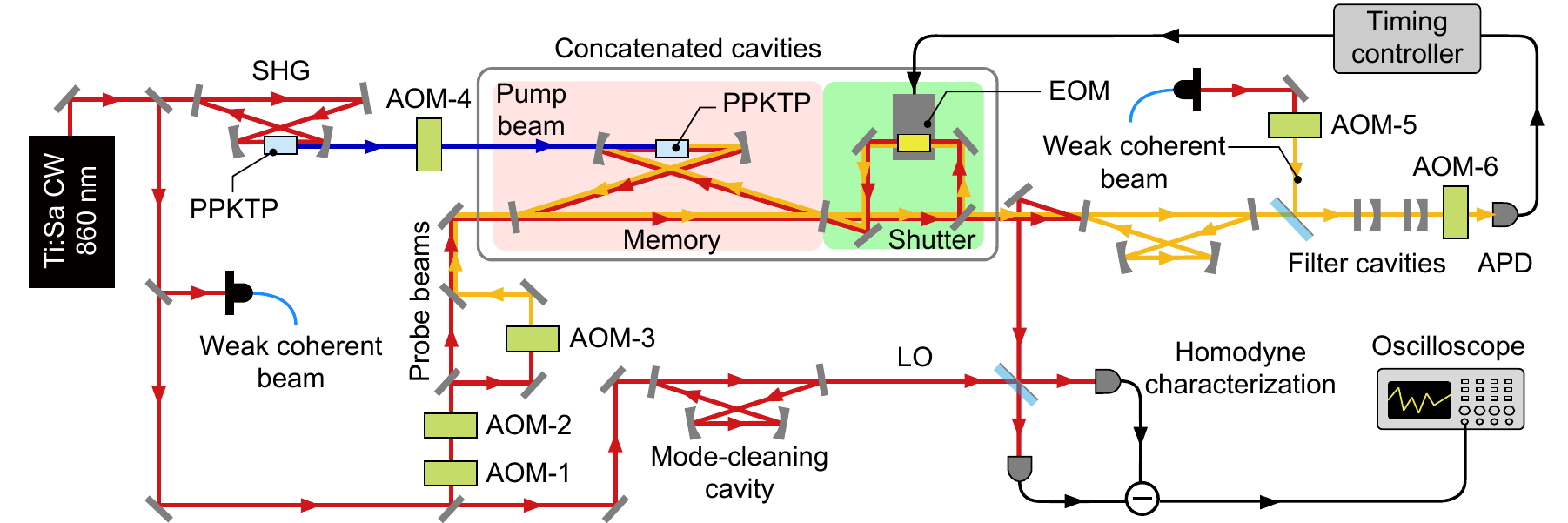}
\caption{Experimental setup.
The light source is a continuous-wave (CW) Ti:sapphire laser operating at 860 nm. 
Optical frequencies of beams are distinguished by colors: 
Red (gray), orange (light gray), and blue (dark gray) stand for signal, idler, and pump frequencies, respectively, while small detuning is not indicated by colors.
Black lines stand for electronic signal lines.
Here SHG denotes a second harmonic generator, PPKTP a periodically-poled KTiOPO$_4$ crystal.
}
\label{fig:setup}
\end{figure*}

Figure~\ref{fig:setup} shows our experimental setup.
This is based on a previous system to store single-photon states \cite{Yoshikawa.PRX2013,Makino.SciAdv2016}.
In order to probe the phase inside the memory cavity, we introduce coherent beams, which we call probe beams, inside the memory cavity for both the signal and idler fields.
These are utilized for feedback control of optical phases outside the concatenated cavities, such as the auxiliary coherent beam which is combined with the idler field to create superpositions, and a local oscillator (LO) beam at the signal frequency for homodyne characterization.
The relative phase between the signal and idler probe beams is also stabilized by monitoring their phase-sensitive parametric amplification.
In addition, the signal probe beam is utilized for precise control of the resonance frequency of the memory cavity.
However, these probe beams should not be present when superposition states are created, because they destroy the heralding and homodyne signals.
Therefore, the probe beams are periodically chopped by using acousto-optic modulators (AOMs), and the superposition states are created when the probe beams are absent. 
We switch the system between the feedback-control phase and the state-creation phase with a period of 200~ms, during which 50~ms is utilized for the state creation.
The duty cycle is smaller than that of the previous experiments without the probe beams \cite{Yoshikawa.PRX2013,Makino.SciAdv2016}, because we have to wait for the decay of the probe beams inside the memory cavity.
The auxiliary coherent beam at the idler frequency determines the created superposition state $\alpha \ket{0} + \beta e^{i\theta} \ket{1}$. 
The amplitudes $\alpha$ and $\beta$ as well as the phase $\theta$ are independently adjustable by changing the amplitude and the phase of the auxiliary coherent beam \cite{Bimbard.NPhoton2010,Yukawa.OptExpress2013,Neergaard.PRL2010}.
An avalanche photodiode (APD) heralds the superposition state, and after a predetermined storage time controlled by a field-programmable gate array, the signal state is released.
A typical photon-detection rate for creation of $(\ket{0} + e^{i\theta}\ket{1})/\sqrt{2}$ is about 1,500 counts per second, among which fake clicks caused by stray light are estimated to be about 50 counts per second.
For simplicity, some beams for controlling the optical systems are omitted from the figure. 
More details of the methods are given in the Supplemental Material (SM) \cite{supplemental}.

One thing to note here is that the resonance frequency of the memory cavity is detuned by about 300 kHz from the frequency of the LO for homodyne characterization.
This detuning is for the purpose of avoiding unexpected photons stored by the memory, which originate from scattering of the LO (probably back-scattering at the photo-diodes of the homodyne detector).
Such scattering rate is very low, but we emphasize that even scattered light at the single-photon level disturbs the experiment if it couples to the memory cavity.
Thanks to the detuning, the effects of the LO scattering are almost entirely removed.
The signal states characterized by the homodyne detection are on a rotational frame and evolve temporally at the angular velocity of about $2 \pi \times 300$ kHz.
This will be seen later in the experimental results as phase shifts of the quantum state during storage.
We also note that this detuning is not necessary if we could use a sufficiently good optical isolator, by which the scattered LO is removed \cite{Morin.OptLett2012}.

\begin{figure}
\centering
\includegraphics[width = 0.45\textwidth, clip]{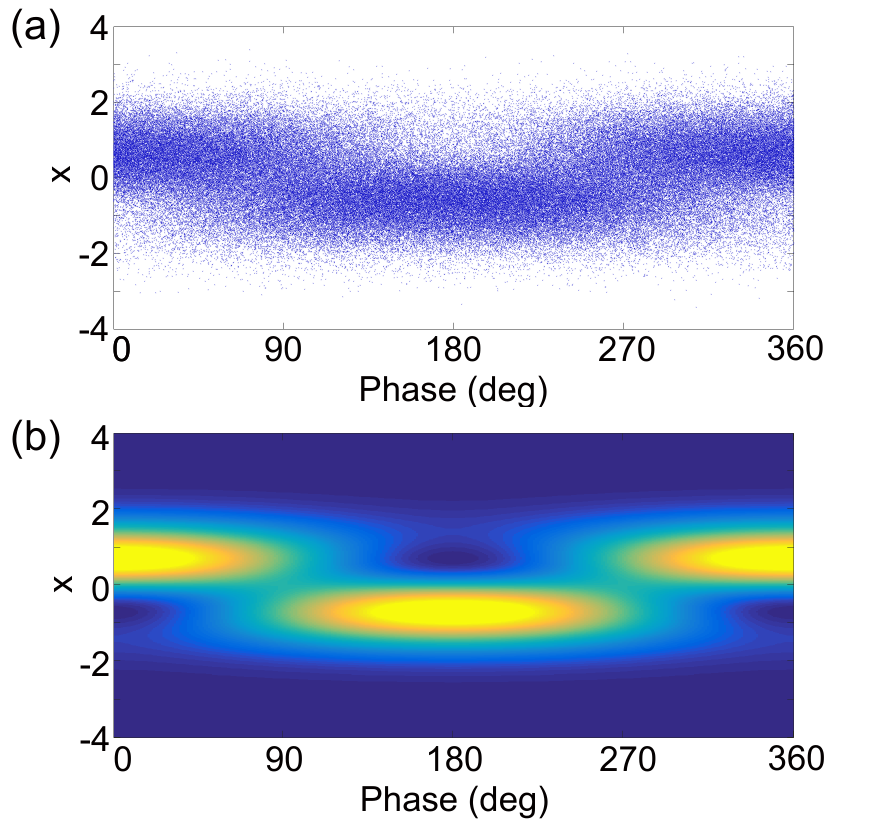}
\caption{Quadrature distributions of $(\ket{0}+\ket{1})/\sqrt{2}$ for various measurement phases. 
(a) Experimental quadrature distribution obtained by homodyne measurements, when crated states are immediately released.
(b) Theoretical quadrature distribution for the ideal pure state.}
\label{fig:quadrature}
\end{figure}

\begin{figure*}
\centering
\includegraphics[width = 0.9\textwidth, clip]{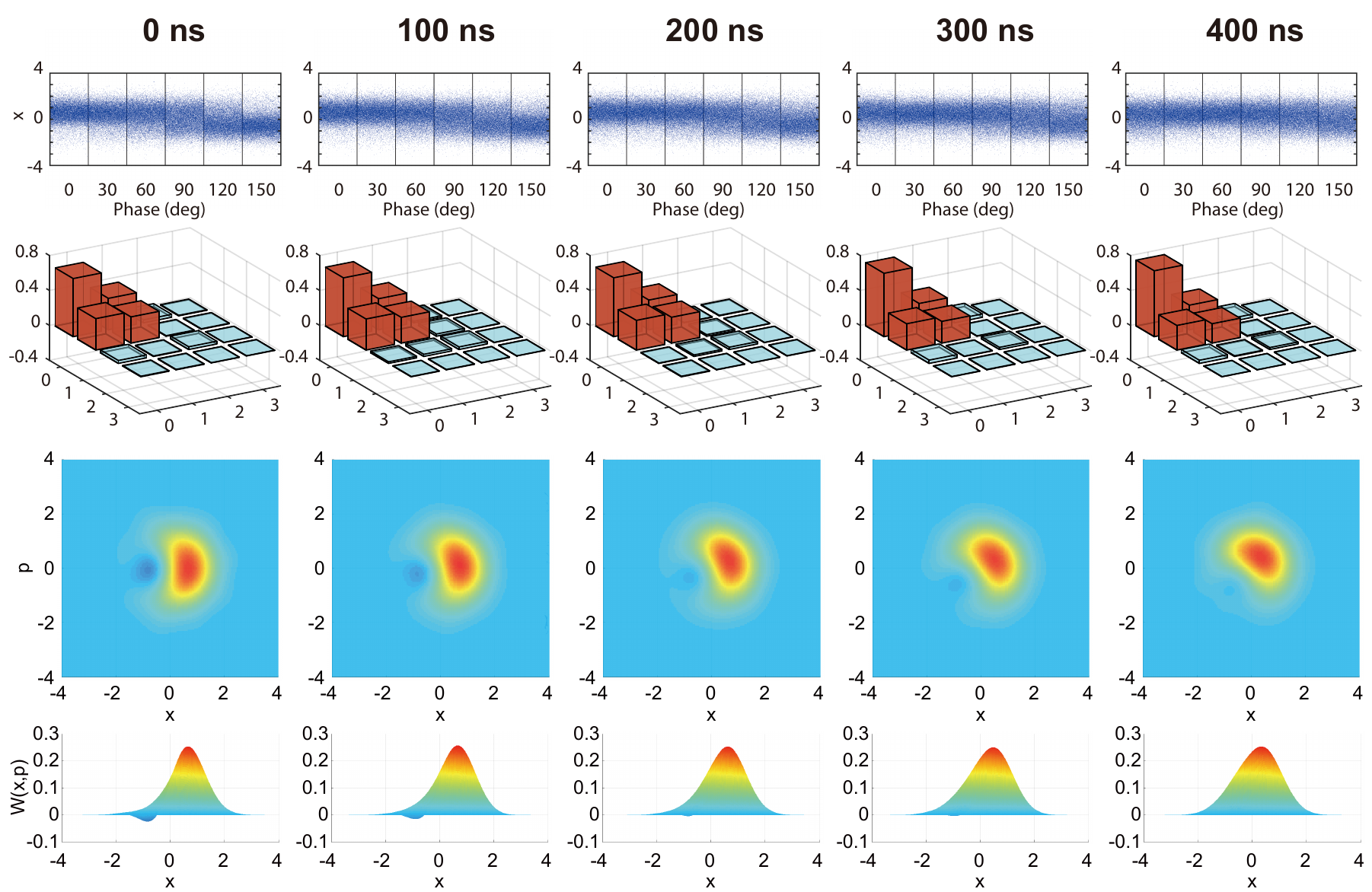}
\caption{Experimental results of storing $(\ket{0}+\ket{1})/\sqrt{2}$ with various storage time.
From left to right, the storage time varies from 0 to 400 ns in steps of 100 ns.
First row: quadrature distributions obtained by homodyne measurements.
Second row: absolute values of reconstructed density matrix elements.
Red elements represent the subspace spanned by $\ket{0}$ and $\ket{1}$, and light blue elements represent the multi-photon components.
Third row: Wigner functions $W(x,p)$ with $\hbar = 1$ seen from the top, corresponding to the reconstructed density matrices.
Fourth row: Wigner functions $W(x,p)$ seen from the side.
Minimum values are $W(-0.84, 0.09) = -0.024$ in 0ns, $W(-0.88, -0.23) = -0.015$ in 100 ns, and $W(-0.84,-0.37)=-0.004$ in 200 ns, where statistical errors of the Wigner values  are estimated as $\pm0.001$}.
\label{fig:storage_0deg_1:1}
\end{figure*}

We first show the phase-sensitive quadrature distribution of $(\ket{0} + e^{i\theta}\ket{1})/\sqrt{2}$ obtained by the homodyne detection in Fig.~\ref{fig:quadrature}.
Figure~\ref{fig:quadrature}(a) shows the experimental quadrature distribution with immediate release after heralding, while Fig.~\ref{fig:quadrature}(b) shows the theoretical distribution with an ideal pure state $(\ket{0} + \ket{1})/\sqrt{2}$.
The experimental quadrature distribution is obtained by repeating the sequence of storage and release 5,000 times at each of the measurement phases from $0^\circ$ to $350^\circ$ at an interval of $10^\circ$.
The horizontal axis corresponds to the measurement phase, while the vertical axis corresponds to the quadrature value obtained from the homodyne signal.
The experimental distribution shows the phase-sensitive nature of the quantum states similar to the theoretically predicted distribution.
This distribution contains sufficient information to reconstruct the density matrix and the Wigner function \cite{Lvovsky.JOptB2004}.

Next we implemented various storage times for $(\ket{0}+\ket{1})/\sqrt{2}$ estimating the released quantum state for each case.
The experimental results are shown in Fig.~\ref{fig:storage_0deg_1:1}.
The storage times are 0 ns, 100 ns, 200 ns, 300 ns, and 400 ns from left to right.

The first row shows the quadrature distributions, where measurement phases are 0$^\circ$, 30$^\circ$, 60$^\circ$, 90$^\circ$, 120$^\circ$, and 150$^\circ$, and the number of points is 20,000 for each measurement phase.
Even for the case of 400 ns storage time, the phase-sensitive nature of the memorized states clearly remains.
The distribution shifts horizontally depending on the storage time, which is explained by the 300 kHz detuning explained above.
From the quadrature distributions, the density matrices and the Wigner functions are estimated.

The second row shows the reconstructed density matrices, where the absolute value is taken for each matrix element.
The matrix elements of the subspace up to one photon are colored in red (gray).
Multi-photon components, colored in light blue (light gray), are very small (less than 5\%).
We can see that the off-diagonal elements are nicely preserved during the storage, though some portion of the single-photon component $\ket{1}\bra{1}$ is converted to a vacuum component $\ket{0}\bra{0}$ mainly due to intracavity losses (about 0.2\% for each round trip of 1.5 m).
The phase shifts of the off-diagonal elements due to the detuning are not visible, because the absolute value is taken for each element.
In the SM, real and imaginary parts of the density matrices are separately plotted, so that one see the phase shifts during the storage \cite{supplemental}.
We also discuss in the SM the amount of phase fluctuations estimated from the density matrices \cite{supplemental}.

The third and fourth rows are the Wigner functions seen from the top and the side.
Seen from the top, the state rotates with an angle proportional to the storage time.
Again, this rotation corresponds to the 300 kHz detuning mentioned above.
Seen from the side, the negative region of Wigner function clearly remains with a storage time of 100 ns, and it is still visible with a storage time of 200 ns.
The negative region is not at the origin, unlike for the previous single-photon case \cite{Yoshikawa.PRX2013}, and therefore this negativity exhibits nonclassicality that cannot be preserved without the phase information of the quantum state.
Moreover, even with longer storage times where the negativity is lost, the Wigner function still satisfies a criterion of genuine non-Gaussianity (i.e., the distribution cannot be produced by classical mixture of Gaussian states) \cite{Genoni.PRA2013}, which we discuss in more detail in the SM \cite{supplemental}.

We also confirmed that the coefficients $\alpha$, $\beta$, $\theta$ of the superposition $\alpha \ket{0} + \beta e^{i\theta} \ket{1}$ are controllable in the state-preparation stage.
Experimental results with other states, such as $(\ket{0} -i\ket{1})/\sqrt{2}$ and $(\ket{0} + \sqrt{2}\ket{1})/\sqrt{3}$, are shown in the SM \cite{supplemental} and these are consistent with the above results.

In conclusion, we experimentally demonstrated creation, storage, and on-demand release of phase-sensitive optical quantum states like $\alpha\ket{0} + \beta e^{i\theta}\ket{1}$ by employing a concatenated cavity system.
We succeeded in storing quantum states with a negative region in the Wigner function shifted away from the phase-space origin --- a feature only possible with a phase-preserving memory system.
Generally, the phase-preserving mechanism is very important for a memory system, because many quantum features such as quantum entanglement are lost without the phase information.
Our demonstration paves the way for future demonstrations of storing various quantum states such as Schr\"odinger cat states, cubic phase states, and so forth.
The event rate will increase if we use superconductive single-photon detectors \cite{LeJeannic.2016OptLett} or photon-number resolving detectors with transition-edge sensor \cite{Fukuda.OptExpress2011}, which have higher quantum efficiencies.

This work was partly supported by JSPS KAKENHI, CREST of JST, and APSA, of Japan.
Y. H. acknowledges support from ALPS. P.v.L. acknowledges support from Q.Link.X (BMBF) in Germany.


\clearpage

\onecolumngrid

\begin{center}
{
\large\bf Supplemental Material for \\ All-Optical Storage of Phase-Sensitive Quantum States of Light}
\vspace{\baselineskip}
\end{center}

\renewcommand{\thefigure}{S\arabic{figure}}
\allowdisplaybreaks[4]

\twocolumngrid

\section{Methods}

One experimental setup is shown in Fig.~\ref{fig:setup} of the main text.
The light source is a continuous-wave Ti:Sapphire laser operating at the wavelength of 860 nm (MBR-110, Coherent).
The second harmonic with the power of about 1.5 mW is used as a pump beam to create photon pairs inside the memory cavity. 
The memory cavity, containing a periodically-poled KTiOPO$_4$ (PPKTP) crystal (Raicol) as a nonlinear optical medium, has a round-trip length of about 1.4 m, while the shutter cavity, containing an RTP EOM (Leysop), has that of about 0.7 m. 
The PPKTP crystal is type-0 quasi-phase-matched and has a length of 10 mm. 
The reflectivity of the coupling mirror between the memory and the shutter is about 98\%, and that of the outcoupling mirror of the shutter is about 72\%. 
The EOM in the shutter is driven by a high-voltage switch (Bergmann Messger\"ate Entwicklung). 
The latency of the high-voltage switch is less than 50 ns, and the applied voltage is estimated as about 900 V.

Three filter cavities are applied to the idler field. 
The first one is bow-tie shaped and has a round-trip length of about 250 mm. 
The signal field is reflected by this first filter cavity and directed to a homodyne detector for the state characterization, while the idler field is transmitted through the cavity. 
Two additional cavities are Fabry-Perot cavities. 
A single tooth of the frequency comb of the memory cavity is selected by these cavities and directed to a photon detector. 
The auxiliary beam to determine the superposition state is combined with the idler field at a beam splitter between the first and the second filter cavities. 
In the following, we call this auxiliary beam a ``displacement beam'' for clarity, because there are also other auxiliary beams in the methods. 
The photon detector contains a silicon avalanche photodiode (SPCM-AQRH-14-FC, Excelitas Technologies) and is coupled with an optical fiber. 
The heralding event rate depends on the amplitude of the displacement beam, and a typical rate for balanced conditions $(\ket{0} + e^{i\theta}\ket{1})/\sqrt{2}$ is about 1,500 counts per second (without compensation of the duty cycle).

Several AOMs are used in the experiment, as depicted in Fig.~\ref{fig:setup}. 
AOM-1 and AOM-2 are for chopping the probe beams. 
Driving signals for these AOMs are on when the optical systems are feedback controlled, and diffracted beams are used as probe beams, while they are off when the superposition states are created. 
AOM-3, AOM-4, AOM-5 are for shifting the laser frequency by the free spectral range of the memory cavity.  
AOM-6 is inserted for protecting the APD from the idler probe beam, by coupling a diffracted beam to the APD and switching the driving signal. 
The detuning of the signal mode from the LO frequency, explained in the main text, is controlled via the difference of the driving frequencies between AOM-1 and AOM-2. 
The signal probe beam is detuned by 200 kHz, and by locking the memory cavity to almost the bottom of the fringe in the transmitted signal probe power, detuning of the signal mode by about 300 kHz from the LO is stably realized. 
Frequencies and phases of the driving signals for the AOMs are precisely controlled by direct digital synthesizers (AD9959, Analog Devices Inc.).

Optical frequency relations are summarized as follows.
Even though several detunings are employed in the experiment, the relation of laser frequencies $\omega_\text{p} = \omega_\text{l} + \omega_\text{d}$ is exact, where $\omega_\text{p}$, $\omega_\text{l}$, and $\omega_\text{d}$ denote the optical frequencies of the pump beam, the LO beam, and the displacement beam to determine the superposition state, respectively.
Experimentally, this means that AOM-4 and AOM-5 are driven by the same electronic signals (about 214 MHz, close to the free spectral range of the memory cavity). 
Thanks to this relation, at the time when the heralding signal is detected, the phases $\theta$ of the created superposition states $\alpha \ket{0} + \beta e^{i\theta} \ket{1}$ inside the memory cavity are the same for all events in the rotating frame of the LO, independent of the state-creation timing, determined solely by the relative phase of the displacement beam.
This situation is the same as for other phase-sensitive heralding experiments without storage \cite{Bimbard.NPhoton2010,Yukawa.OptExpress2013,Neergaard.PRL2010}.
Starting from the same superposition phases, the memory-cavity mode detuned by 300 kHz from the LO rotates the stored states by an angle proportional to the storage time.

Figure~\ref{fig:setup} is somewhat simplified from the actual setup for simplicity. 
From Fig.~\ref{fig:setup}, some beams for controlling the optical systems are omitted, which we call ``lock beams'' in the following. 
The lock beams are periodically chopped in the same way as the probe beams. 
For example, for each cavity, a lock beam is employed for the purpose of obtaining an error signal for locking the cavity length. 
As for the memory cavity, first a lock beam is used for roughly locking the cavity length, and after the rough locking, next the transmitted signal probe beam is used for precisely locking the detuning frequency. 
The lock beam is injected into the memory cavity with vertical polarization, while the pump, signal and idler fields are in horizontal polarization. 
Another lock beam is for locking the displacement beam to the idler probe beam. 
Since the displacement beam is very weak, it is very hard to directly use its interference fringe for locking.  
Therefore, the displacement beam is first locked to a stronger lock beam, and then these are attenuated and locked to the idler probe beam.

\section{Storage of single-photon states}

As a preliminary experiment, we tested storage and release of single-photon states.
The displacement beam is blocked, while the other experimental elements are the same as for the experiment with superposition states.
This part is almost the same as our former experiments \cite{Yoshikawa.PRX2013,Makino.SciAdv2016}, but the difference is that now the probe beams are introduced by which a phase-sensitive full quantum tomography is conducted, whereas in the former experiments phase insensitivity of the quantum states was assumed. 

This preliminary experiment is performed in order to obtain wave-packet envelope functions $\Psi(t)$ of the released quantum states for various storage times by the principal component analysis \cite{MacRae.PRL2012,Morin.PRL2013,Yoshikawa.PRX2013}. 
Quadrature values of the target quantum states are obtained from continuous homodyne signals, denoted by $\hat{x}(t)$, via weighted integration $\hat{x}=\int dt\Psi(t)\hat{x}(t)$. 
The functions $\Psi(t)$ obtained in this preliminary experiment are also used for the characterization of superposition states.

\begin{figure}[tbp]
\centering
\includegraphics[scale = 0.75]{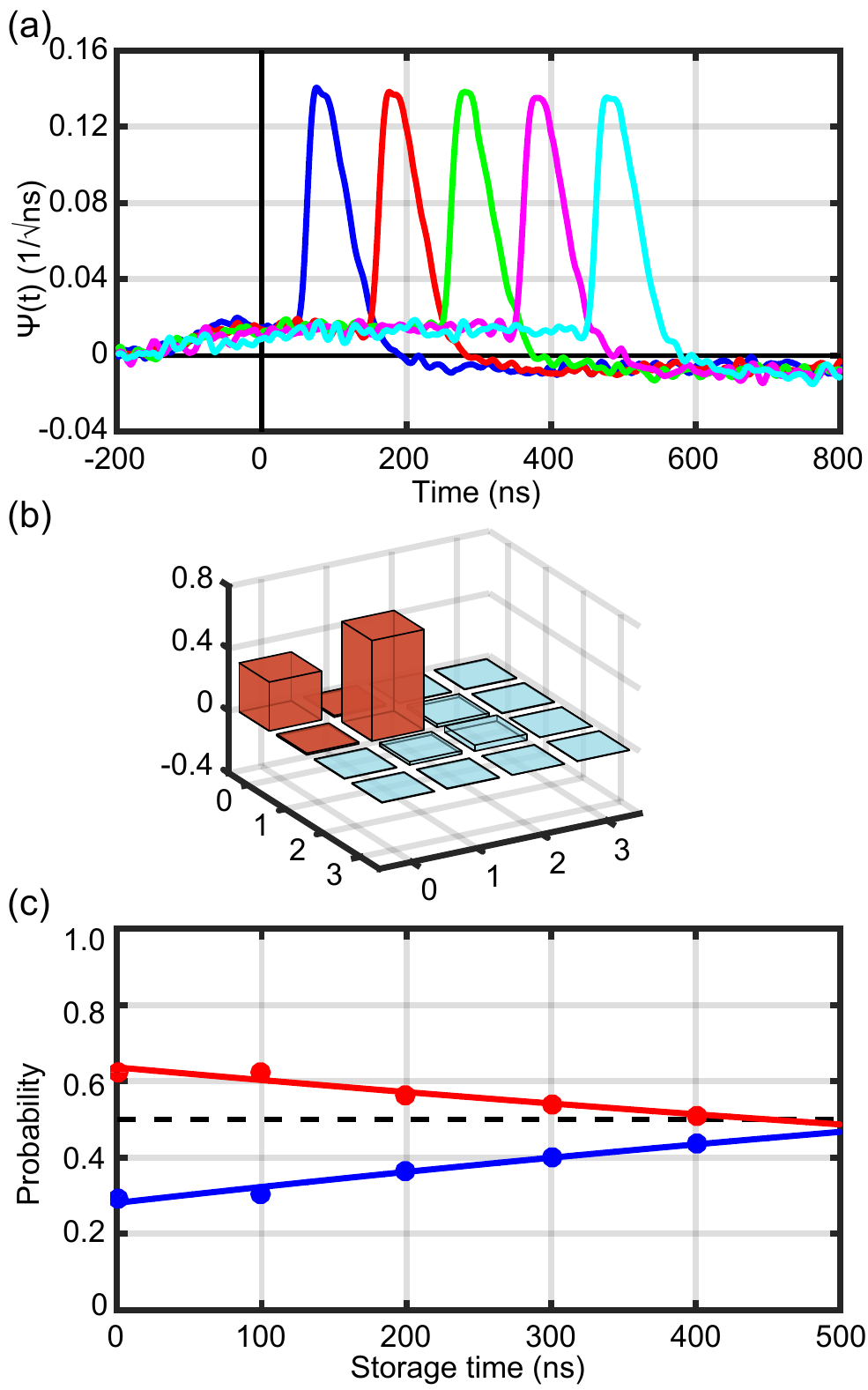}
\caption{Storage and release of single-photon states.
(a)~Wave-packet envelope functions $\Psi(t)$ of the released states. 
Storage times are 0 ns (blue), 100 ns (red), 200 ns (green), 300 ns (magenta), and 400 ns (cyan). 
The time origin corresponds to timings of heralding signals.
(b)~Reconstructed density matrix with 0 ns storage time (immediate release), where the absolute value is taken for each matrix element.
(c)~Single-photon fraction and vacuum fraction for various storage times.
Red: single-photon fraction. Blue: vacuum fraction. 
Circles show experimental values, whereas traces are fitted curves assuming exponential decay.}
\label{fig:preliminary}
\end{figure}

The results of the preliminary experiment are shown in Fig.~\ref{fig:preliminary}.
Figure~\ref{fig:preliminary}(a) shows estimated wave-packet envelope functions $\Psi(t)$ for storage times of 0 ns, 100 ns, 200 ns, 300 ns, and 400 ns. 
The horizontal axis is relative time where the timing of the heralding signal corresponds to 0 ns. 
The shape of the wave packet is independent of the storage time, except for small leakage before release, and appropriately shifted in accordance with the storage time. 
Figure~\ref{fig:preliminary}(b) shows the reconstructed density matrix of a single-photon state immediately released after heralding.
Here, the absolute value is taken for each matrix element. 
Off-diagonal elements are negligibly small, clearly showing phase-insensitivity of the heralded single-photon state.  
Figure~\ref{fig:preliminary}(c) shows the single-photon fraction and the vacuum fraction for various storage times. 
The single-photon fraction exponentially decays, due to small optical losses, and the estimated half-life time is 1.3 $\mu$s.

\section{Storage of superposition states with various coefficients}

\begin{figure*}[tbp]
\centering
\includegraphics[width = 0.9\textwidth]{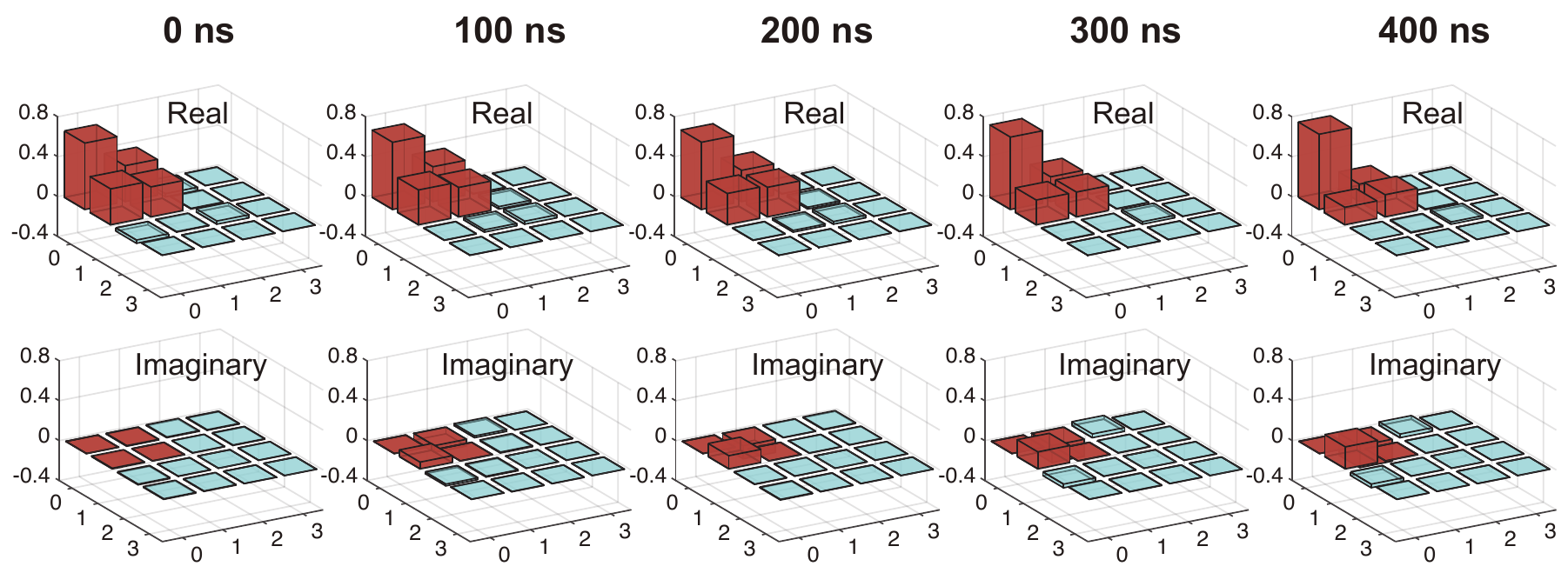}
\caption{
Real and imaginary parts of the reconstructed density matrices in Fig.~\ref{fig:storage_0deg_1:1} in the main text, storing $(\ket{0} + \ket{1})/\sqrt{2}$.
Upper row: real parts of the density matrices.
Lower row: imaginary parts of the density matrices.}
\label{fig:densmat_0deg_1:1}
\end{figure*}

\begin{figure*}
\includegraphics[width = 0.9\textwidth]{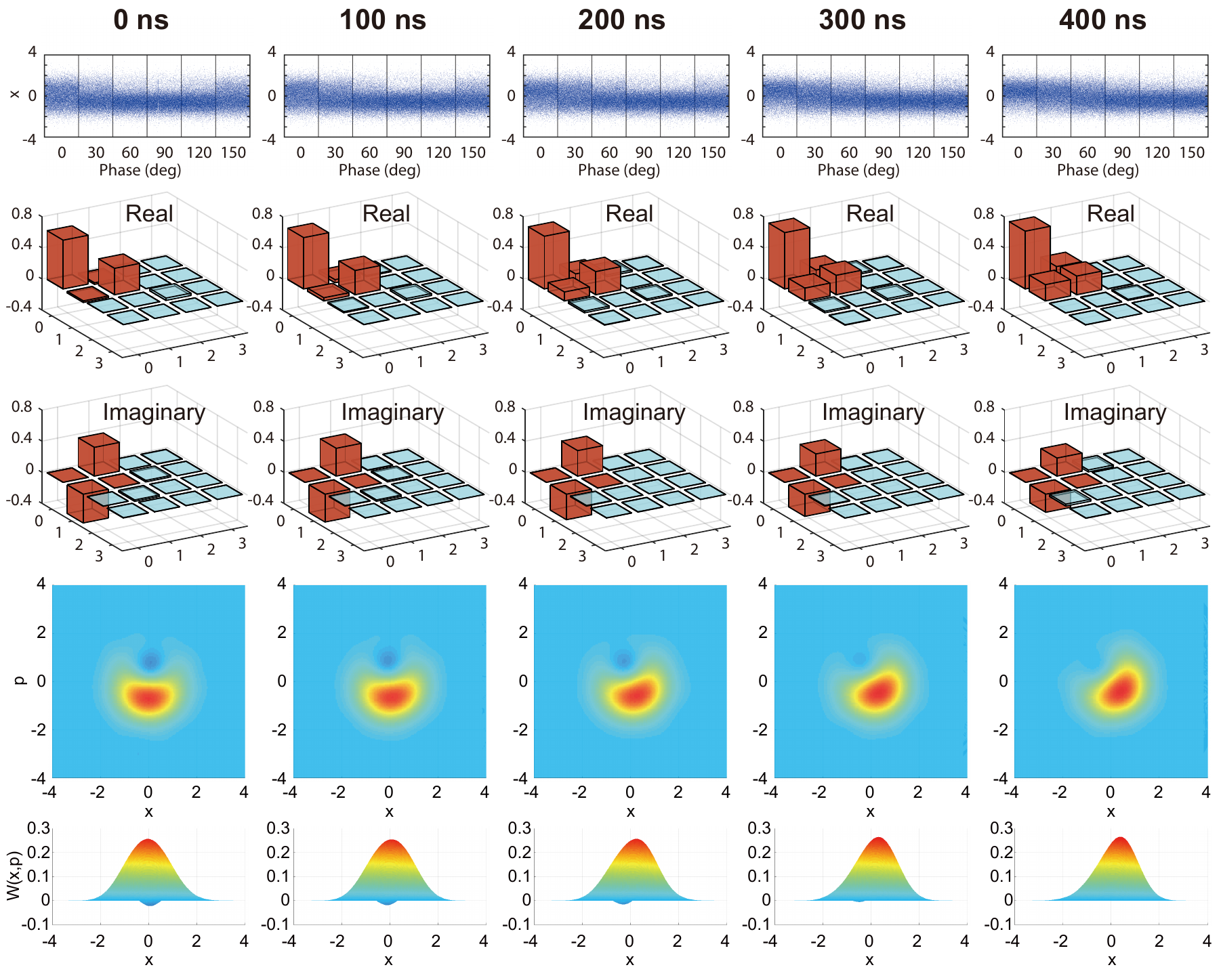}
\caption{
Experimental results of storing $(\ket{0}-i\ket{1})/\sqrt{2}$.
From left to right, the storage time varies from 0 to 400 ns.
First row: quadrature distributions.
Second row: real parts of reconstructed density matrices.
Third row: imaginary parts of reconstructed density matrices.
Fourth row: Wigner functions seen from the top.
Fifth row: Wigner functions seen from the side.
Minimum values are $W(0.06, 0.75) = -0.022$ in 0 ns, $W(-0.10, -0.85) = -0.019$ in 100 ns, and $W(-0.29, 0.81) = -0.016$ in 200 ns, where statistical errors of the Wigner values are estimated as $\pm0.001$.
}
\label{fig:storage_90deg_1:1}
\end{figure*}

\begin{figure*}
\includegraphics[width = 0.9\textwidth]{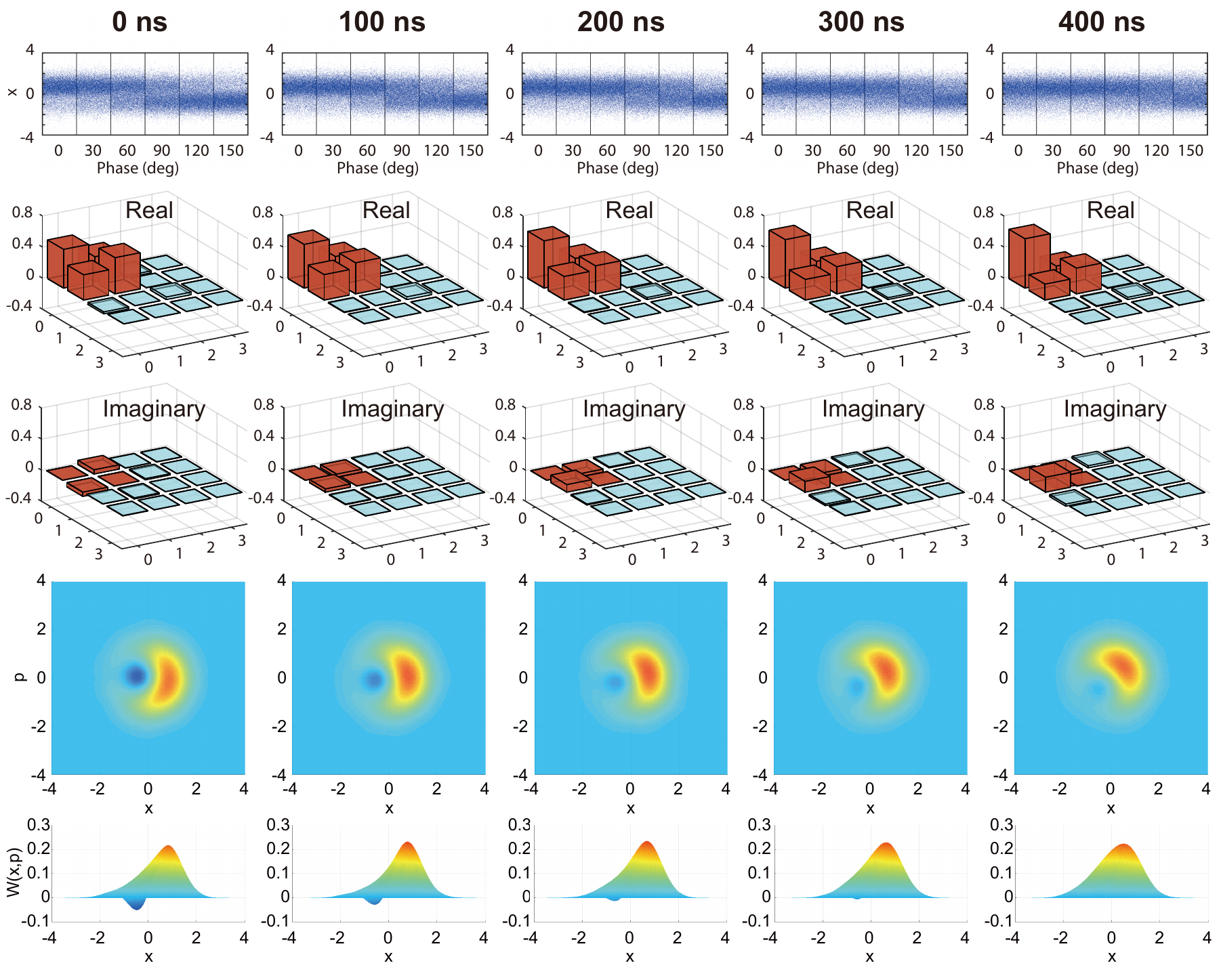}
\caption{Experimental results of storing $(\ket{0} +\sqrt{2} \ket{1})/\sqrt{3}$.
From left to right, the storage time varies from 0 to 400 ns.
First row: quadrature distributions.
Second row: real parts of reconstructed density matrices.
Third row: imaginary parts of reconstructed density matrices.
Fourth row: Wigner functions seen from the top.
Fifth row: Wigner functions seen from the side.
Minimum values are $W(-0.47, 0.06) = -0.051$ in 0 ns, $W(-0.58, -0.11) = -0.029$ in 100 ns, and $W(-0.64, -0.20) = -0.013$ in 200 ns, where statistical errors of the Wigner values are estimated as $\pm0.001$.
}
\label{fig:storage_0deg_1:2}
\end{figure*}

Here, we present experimental results for various coefficients of the superposition states.

Figure~\ref{fig:densmat_0deg_1:1} shows the same results as Fig.~\ref{fig:storage_0deg_1:1} in the main text, storing $(\ket{0}+\ket{1})/\sqrt{2}$, but the real and imaginary parts of the density matrices are separately shown. 
Due to the phase rotation caused by detuning of the memory cavity, the imaginary parts of the off-diagonal elements $\ket{0}\bra{1}$ and $\ket{1}\bra{0}$ gradually emerge when the storage time becomes longer.

Figure~\ref{fig:storage_90deg_1:1} shows the experimental results of storing $(\ket{0}-i\ket{1})/\sqrt{2}$, which is created by shifting the phase of the displacement beam compared to the case of $(\ket{0}+\ket{1})/\sqrt{2}$.
As for the density matrices, in comparison with the case of $(\ket{0}+\ket{1})/\sqrt{2}$, the diagonal elements are almost unchanged, while the phases of the off-diagonal elements  are shifted by $90^\circ$. 
Correspondingly, the Wigner functions are rotated by $90^\circ$.

Figure~\ref{fig:storage_0deg_1:2} shows the experimental results of storing $(\ket{0} + \sqrt{2}\ket{1})/\sqrt{3}$, which is created by attenuating the displacement beam.
The $\ket{1}\bra{1}$ components in the density matrices are larger, resulting in Wigner functions with larger negative values at positions closer to the origin compared to the results above.

The above results show the controllability of both the phase and amplitude of the superposition in the creation stage. 
Our memory system is capable of storing such a variety of quantum states.
A rotation of quantum states during storage is consistent with the detuning explained in the main text and in the methods.

\section{Quantum non-Gaussianity}

Here we discuss the genuine non-Gaussianity of the released states. 
Even though the negative region in the experimental Wigner functions is destroyed at a storage time of 400 ns, we will see that a quantum-mechanically meaningful non-Gaussianity still remains. 

A well-known fact is that, for a pure state, any non-Gaussian state has some negative region in its Wigner function. 
On the other hand, for a mixed state, the situation is more involved. 
A positive non-Gaussian Wigner function may or may not be a statistical mixture of Gaussian Wigner functions. 
A non-Gaussian Wigner function that cannot be expressed as a statistical mixture of Gaussian Wigner functions is referred to as genuine non-Gaussian, or quantum non-Gaussian. 
A negative region in the Wigner function is apparently a sufficient condition for quantum non-Gaussianity, but not a necessary condition. 

Although it is generally a difficult task to derive a necessary and sufficient condition for quantum non-Gaussianity, fortunately there are known sufficient conditions more relaxed than a negative region in the Wigner function. 
In particular, we employ a criterion based on the Wigner function value at the origin and the mean photon number \cite{Genoni.PRA2013}. 
If a density matrix $\hat{\rho}$ of a single harmonic oscillator represents a statistical mixture of Gaussian states, the following $\Delta(\hat{\rho})$ is always nonnegative: 
\begin{equation}
\Delta[\hat{\rho}] = W[\hat{\rho}](0,0) - \frac{1}{\pi}\exp\{-2n[\hat{\rho}](n[\hat{\rho}]+1)\}, 
\label{eq:nonGauss}
\end{equation}
where $W[\hat{\rho}](x,p)$ is the Wigner function for $\hat{\rho}$, and $n[\hat{\rho}]\coloneqq\Tr(\hat{\rho}\hat{a}^\dagger\hat{a})$ represents the mean photon number of $\hat{\rho}$. 
Here, $\hat{a}$ and $\hat{a}^\dagger$ are annihilation and creation operators as usual, and the photon number operator is $\hat{n}=\hat{a}^\dagger\hat{a}$. 

The derivation of Eq.~\eqref{eq:nonGauss} is summarized as follows \cite{Genoni.PRA2013}. 
First, we consider pure Gaussian states $\hat{\rho}=\ket{\psi}\bra{\psi}$, and we find that the minimum value of the Wigner function origin for a given mean photon number $\bar{n}$ is $(1/\pi)\exp[-2\bar{n}(\bar{n}+1)]$. 
Next, using downward convexity of the function $\xi(n)\coloneqq(1/\pi)\exp[-2n(n+1)]$, and using linearity of the Wigner function $W[\sum_kp_k\hat{\rho}_k](x,p)=\sum_kp_kW[\hat{\rho}_k](x,p)$ and that of the mean photon number $n[\sum_kp_k\hat{\rho}_k]=\sum_kp_kn[\hat{\rho}_k]$, we find, for any statistical mixture of Gaussian states $\hat{\rho}=\sum_kp_k\ket{\psi_k}\bra{\psi_k}$,
\begin{align}
W[\hat{\rho}](0,0)
&=\sum_kp_kW\bigl[\ket{\psi_k}\bra{\psi_k}\bigr](0,0) \notag\\
&\ge\sum_kp_k\xi\bigl(n\bigl[\ket{\psi_k}\bra{\psi_k}\bigr]\bigr) \notag\\
&\ge\xi\bigl(\sum_kp_kn\bigl[\ket{\psi_k}\bra{\psi_k}\bigr]\bigr) =\xi(n[\hat{\rho}]).
\end{align}
We note that, since both the Wigner origin value $W[\hat{\rho}](0,0)$ and the mean photon number $n[\hat{\rho}]$ are invariant under phase-space rotation operations, the above criterion itself is only relevant to diagonal elements of a density matrix in the number basis. 
In particular, the Wigner origin value is determined by the difference between even- and odd-photon-number contributions $W[\hat{\rho}](0,0)=(1/\pi)\sum_m(\rho_{2m,2m}-\rho_{2m+1,2m+1})$. 
However, as we will discuss next, the criterion can be optimized for individual states by Gaussian operations, thereby quantum non-Gaussianity of some phase-sensitive states being witnessed.

Since Gaussian unitary operations form a group transitively acting on pure Gaussian states, quantum non-Gaussianity is a property invariant under Gaussian unitary operations. 
Therefore, Eq.~\eqref{eq:nonGauss} can be optimized by Gaussian operations for the witness of quantum non-Gaussianity. 
That is, if $\Delta(\hat{U}_\text{G}\hat{\rho}\hat{U}_\text{G}^\dagger)$ is negative for some Gaussian unitary operation $\hat{U}_\text{G}$, the tested quantum state $\hat{\rho}$ is quantum non-Gaussian \cite{Genoni.PRA2013}.

\begin{figure}[tb]
\centering
\includegraphics[scale=0.75]{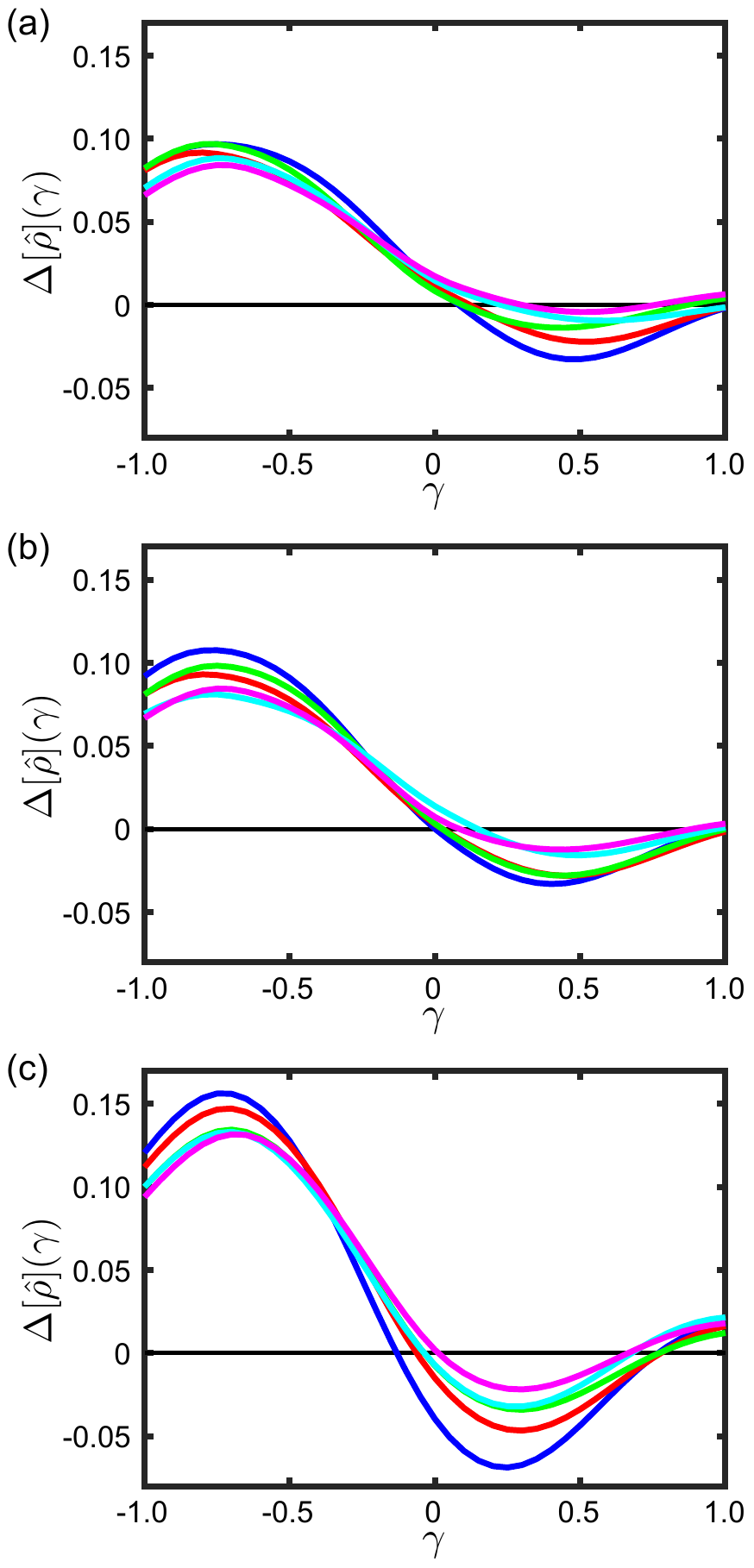}
\caption{
Quantum non-Gaussianity witness $\Delta[\hat{\rho}](\gamma)$. 
(a) Storage of $(\ket{0}+\ket{1})/\sqrt{2}$. 
(b) Storage of $(\ket{0}-i\ket{1})/\sqrt{2}$.
(c) Storage of $(\ket{0}+\sqrt{2}\ket{1})/\sqrt{3}$. 
Storage times are 0 ns (blue), 100 ns (red), 200 ns (green), 300 ns (cyan), and 400 ns (magenta), respectively. 
}
\label{fig:nonGauss}
\end{figure}

In particular, we consider Gaussian corrections by a displacement operation $\hat{D}(\gamma e^{i\phi}) \coloneqq \exp(\gamma e^{i\phi}\hat{a}^\dagger-\gamma e^{-i\phi}\hat{a})$ followed by a squeezing operation $\hat{S}(\zeta e^{i\varphi}) \coloneqq \exp[(\zeta e^{i\varphi}/2)\hat{a}^\dagger\hat{a}^\dagger - (\zeta e^{-i\varphi}/2)\hat{a}\hat{a}]$. 
Intuitively, the displacement operation shifts the Wigner function so that the value at the origin becomes smaller, and the following squeezing operation, which does not change the value at the origin, minimizes the mean photon number. 
Therefore, we set the direction of the displacement, expressed by $\phi$, to the direction of the dip in the Wigner function, and the axis of the squeezing or antisqueezing, expressed by $\varphi/2$, to be parallel to the displacement direction from the symmetry, i.e., $\varphi=2\phi$. 
The squeezing degree $\zeta$ is optimized for each displacement $\gamma$. 
To sum up, we define the Gaussian-corrected quantum non-Gaussianity witness $\Delta[\hat{\rho}](\gamma)$ with a parameter $\gamma$ as, 
\begin{align}
\Delta[\hat{\rho}](\gamma)&\coloneqq \Delta\big[\hat{U}_\text{G}[\hat{\rho}](\gamma)\hat{\rho}\hat{U}_\text{G}^\dagger[\hat{\rho}](\gamma)\big], \\
\hat{U}_\text{G}[\hat{\rho}](\gamma)&\coloneqq\hat{S}\big(\zeta(\gamma) e^{2i\phi[\hat{\rho}]}\big)\hat{D}\big(\gamma e^{i\phi[\hat{\rho}]}\big). 
\end{align}

Figure~\ref{fig:nonGauss} shows the quantum non-Gaussianity witness $\Delta[\hat{\rho}](\gamma)$ for the storage of $(\ket{0}+\ket{1})/\sqrt{2}$, $(\ket{0}-i\ket{1})/\sqrt{2}$, and $(\ket{0}+\sqrt{2}\ket{1})/\sqrt{3}$, corresponding to the results in Figs.~4(or \ref{fig:densmat_0deg_1:1}), \ref{fig:storage_90deg_1:1} and \ref{fig:storage_0deg_1:2}, respectively. 
Note that the direction of the displacement $\phi[\hat{\rho}]$ is different for each storage time because the stored state rotates during storage. 
We can see that every trace enters the negative region at some $\gamma$, proving quantum non-Gaussianity up to a storage time of 400 ns. 
We note that the optimal $\gamma$ for $(\ket{0}+\sqrt{2}\ket{1})/\sqrt{3}$ is closer to zero than that for the other two states. 
This is consistent with the position of the negative dip in the Wigner function closer to the origin.

\section{Phase-fluctuation analysis}

From the experimental results, we have seen that the phase-preserving feature of our memory system is in the quantum regime, enough to preserve a negative region of a Wigner function shifted away from the origin, produced by off-diagonal elements of a density matrix. 
Here, we make an estimation of the amount of phase fluctuations in our system from the reconstructed density matrices.

\begin{figure}[tb]
\centering
\includegraphics[scale = 1]{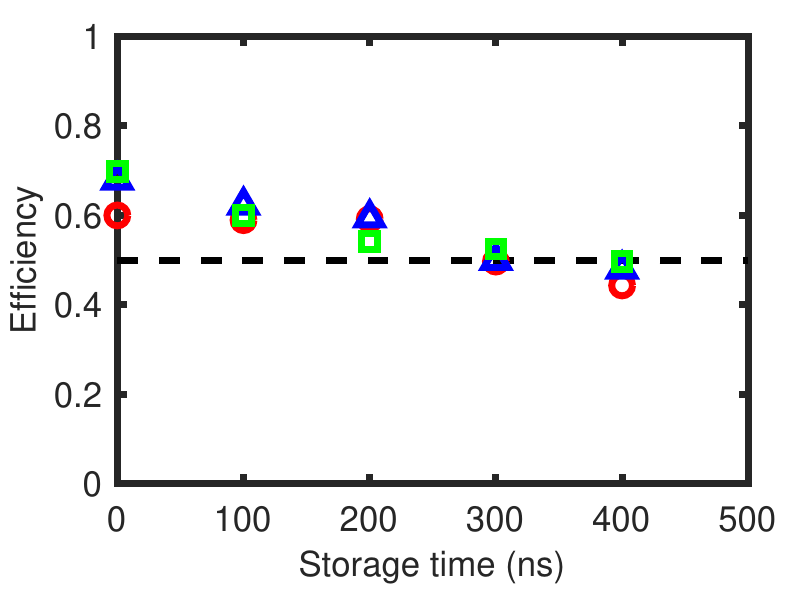}
\caption{
Photon-preserving efficiencies $1-L$ estimated from experimental density matrices for various superposition states and various storage times. 
Red circle: $(\ket{0}+\ket{1})/\sqrt{2}$. Blue triangle: $(\ket{0}-i\ket{1})/\sqrt{2}$. Green square: $(\ket{0}+\sqrt{2}\ket{1})/\sqrt{3}$. 
}
\label{fig:efficiency}
\end{figure}

\begin{figure}[tb]
\centering
\includegraphics[scale = 1]{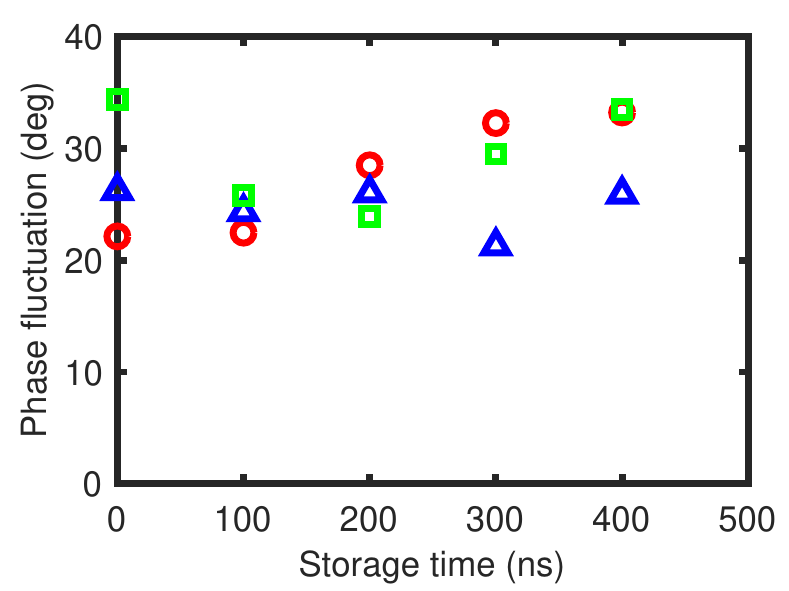}
\caption{
Standard deviations of phase fluctuations $\sigma$ estimated from experimental density matrices for various superposition states and various storage times. 
Red circle: $(\ket{0}+\ket{1})/\sqrt{2}$. Blue triangle: $(\ket{0}-i\ket{1})/\sqrt{2}$. Green square: $(\ket{0}+\sqrt{2}\ket{1})/\sqrt{3}$. 
}
\label{fig:dephasing}
\end{figure}

The reconstructed density matrices deviate from the ideal ones. 
Here we separate the deviations into optical losses (or amplitude damping) and phase fluctuations (or dephasing). 
We consider a density matrix $\hat{\rho}=\sum_{m,n\in\{0,1\}}\rho_{m,n}\ket{m}\bra{n}$ in the number basis up to one photon. 
An optical loss $L$ changes the density matrix as
\begin{align}
\begin{pmatrix}
\rho_{0,0} & \rho_{0,1} \\
\rho_{1,0} & \rho_{1,1}
\end{pmatrix}
\to
\begin{pmatrix}
(1+L)\rho_{0,0} & \sqrt{1-L}\rho_{0,1} \\
\sqrt{1-L}\rho_{1,0} & (1-L)\rho_{1,1}
\end{pmatrix}.
\end{align}
Therefore, the decay in $\rho_{1,0}$ or $\rho_{0,1}$ is the square root of that in $\rho_{1,1}$. 
On the other hand, phase fluctuations do not change the diagonal elements but diminish the off-diagonal elements. 
Assuming a phase fluctuation in a Gaussian distribution $(1/\sqrt{2\pi\sigma^2})e^{-\theta^2/(2\sigma^2)}$ with a standard deviation $\sigma$, the phase fluctuation changes the density matrix as
\begin{align}
\begin{pmatrix}
\rho_{0,0} & \rho_{0,1} \\
\rho_{1,0} & \rho_{1,1}
\end{pmatrix}
\to
\begin{pmatrix}
\rho_{0,0} & e^{-\frac{\sigma^2}{2}}\rho_{0,1} \\
e^{-\frac{\sigma^2}{2}}\rho_{1,0} & \rho_{1,1}
\end{pmatrix}.
\end{align}
This is derived from the integral $\int e^{i\theta}e^{-\theta^2/(2\sigma^2)}d\theta = \int e^{-(\theta-i\sigma^2)^2/(2\sigma^2)}e^{-\sigma^2/2}d\theta = \sqrt{2\pi\sigma^2}e^{-\sigma^2/2}$. 
Precisely speaking, there are other types of experimental imperfections. 
For example, fake clicks in the heralding detector produce a statistical mixture of heralded states and vacuum states $\eta\hat{\rho}+(1-\eta)\ket{0}\bra{0}$. 
The effects of fake clicks are equivalent to those of losses for the case of single-photon states, but different for the case of superposition states. 
Moreover, taking into account multiphoton contributions further complicates the situation. 
However, here we neglect such imperfections and simplify the situation by just considering the two factors, and taking the subspace of the density matrix up to one photon.

Our assumption in this analysis is that the initial quantum states are pure superposition states $\alpha\ket{0}+\beta e^{i\theta}\ket{1}$, where $\alpha$ and $\beta$ are ideally prepared, i.e. we assume $\alpha^2=1/2$, $\beta^2=1/2$ for the experiments of storing $(\ket{0}+\ket{1})/\sqrt{2}$ (Figs.~4, \ref{fig:densmat_0deg_1:1}) or $(\ket{0}-i\ket{1})/\sqrt{2}$ (Fig.~\ref{fig:storage_90deg_1:1}), and $\alpha^2=1/3$, $\beta^2=2/3$ for the experiment of storing $(\ket{0}+\sqrt{2}\ket{1})/\sqrt{3}$ (Fig.~\ref{fig:storage_0deg_1:2}).
Then, we calculate the amounts of losses $L$ and phase fluctuations $\sigma$ consistent with the experimental density matrix $\hat{\rho}$, where the subspace of up to one photon is taken and renormalized (the neglected multiphoton components are about 3\%). 
The single-photon component gives information regarding losses $L$ by the relation 
\begin{align}
1-L=\frac{\rho_{1,1}}{\beta^2}.
\end{align}
On the other hand, the off-diagonal elements, together with the obtained losses, gives information on the phase fluctuations $\sigma$ by the relation
\begin{align}
e^{-\frac{\sigma^2}{2}}=\frac{\lvert\rho_{0,1}\rvert}{\alpha\beta\sqrt{1-L}}.
\end{align}

The calculated photon-preserving efficiencies $1-L$ and phase fluctuations $\sigma$ are shown in Fig.~\ref{fig:efficiency} and Fig.~\ref{fig:dephasing}, respectively. 
Efficiencies in Fig.~\ref{fig:efficiency} are almost consistent with the results of single-photon storage in Fig.~\ref{fig:preliminary}(c). 
As for the phase fluctuations in Fig.~\ref{fig:dephasing}, the results are from 20$^\circ$ to 35$^\circ$, but we could not find a clear tendency like worsening dephasing during storage. 
It seems that other accidental factors obscure the accumulating dephasing during storage. 
The calculated phase fluctuations may look a bit large, but we note that amplitude fluctuations may be converted to phase fluctuations in this analysis. 
If the effective pump power fluctuates, the single-photon fraction of the heralded state fluctuates, making a statistical mixture which has smaller off-diagonal elements, resulting in larger $\sigma$. 
We are aware that fluctuations in the heralding event rate during experiments are much larger than those expected from the Poisson distribution. 
In that sense, the calculated values are considered as upper limits of the actual phase fluctuations in our system. 
Moreover, even for the actual phase fluctuations, we are not sure about whether the phase-instability is caused by the memory cavity itself or by the laser system supplying the LO, the pump, and the displacement beam.

\end{document}